\documentclass[prd,amsmath,amssymb]{revtex4}
 \usepackage{graphicx}
 \usepackage{color}
 \usepackage[colorlinks=true,linkcolor=red]{hyperref}
 \usepackage{hyperref}

\begin{document}

 \title{On avoiding cosmological oscillating behavior for
       $S$-brane solutions with diagonal metrics}

 \author{V. D. Ivashchuk}  
 \email{rusgs@phys.msu.ru}
 \author{V. N. Melnikov}
 \email{melnikov@phys.msu.ru}
 \affiliation{Center for Gravitation and Fundamental Metrology, VNIIMS,
 Institute of Gravitation and Cosmology, 
 Peoples' Friendship University of Russia,
 46 Ozernaya Str., G-361,  Moscow, 119361, Russia}

 \author{D. Singleton}
 \email{dougs@csufresno.edu}
 \affiliation{Department of Physics, California State University, Fresno \\
 Fresno, CA 93740-8031 \\
 and \\
 Escuela de Fisica, Universidad de Costa Rica
 San Jose, Costa Rica}

 \date{\today}

  \begin{abstract}
 In certain string inspired higher dimensional cosmological models
 it has been conjectured that there is generic, 
 chaotic oscillating behavior 
 near the initial singularity -- the Kasner parameters which
 characterize the asymptotic form of the metric ``jump" between
 different, locally constant values and exhibit a never-ending
 oscillation as one approaches the singularity. In this paper we
 investigate a class of cosmological solutions
 with form fields and diagonal metrics which have a ``maximal"
 number of composite electric $S$-branes. We look at two explicit
 examples in $D=4$ and $D=5$ dimensions that do not have chaotic
 oscillating behavior near the singularity. When the composite
 branes are replaced by non-composite branes chaotic oscillating
 behavior again occurs. 
  \end{abstract}


\maketitle

  %%%%%%%%%%%%%%%%%%%%%%%%%%%%%%%%%%%%%%%%%%%%%%%%%%%%%%%%%%%%%%%%%%%%%%%%
  \section{Introduction}

  %%%%%%%%%%%%%%%%%%%%%%%%%%%%%%%%%%%%%%%%%%%%%%%%%%%%%%%%%%%%%%%%%%%%%%%%

 In \cite{IKM1,IKM,IMb0,IM-izh,IMb1, DIM-bil} the behavior of certain
 multidimensional solutions with block-diagonal metrics were
 examined in terms of the billiard representation in generalized
 Lobachevsky spaces. The billiard representation is a graphical
 method of studying the asymptotical motion of the point
 described by the anisotropic minisuperspace coordinates, $z^a (t)$,
 in a region enclosed by ``hard walls"  which come from
 the matter sources. The point with
 coordinates, $z^a (t)$, moves inside this region until it
 encounters a ``wall" at which point it recoils according to the
 reflection law moving toward another ``wall" in the
 manner of a billiard ball on a billiard table.  It was shown that
 certain multidimensional cosmological solutions exhibit
 oscillating behavior near the singularity in the sense
 that the anisotropic part of minisuperspace coordinate functions,
 that characterize the solutions, exhibit a never-ending
 oscillation as one approaches the singularity. The first example
 of such behavior near the singularity was discovered in a
 4-dimensional  Bianchi-IX cosmological, ``mixmaster" model \cite{Mis1} 
 by Belinskii,  Lifshitz  and Khalatnikov in \cite{BKL}.
 
 The ``billiard" description of this behavior in the 4D case
 was suggested by Chitre in \cite{Ch}. More recent
 publications can be found in \cite{Pull,Kir1,Kir2,Mis2}.
 In  \cite{IKM1,IKM,IMb0,KM} this billiard approach of
 Chitre was generalized to the multidimensional case.
 
In the case of multidimensional gravity with either ``perfect-fluid"
or scalar and form fields, one can give conditions under which one does
or does not get never-ending oscillating behavior near the singularity.
Here we give a brief summary of these conditions. 
In \cite{IMb0} a detailed description of the billiard
approach for the  multidimensional model with an
$m$-component ``perfect-fluid" matter source was given.
This model was defined on the manifold
   \begin{equation} \label{1.1}
    M = {\bf R} \times M_{1} \times \ldots \times M_{n},
   \end{equation}
   with the block-diagonal metric described by the line 
   element
   \begin{equation} \label{1.2}
  ds^2 = - \exp[2{\gamma}(t)] dt^2 + \sum_{i=1}^{n}
  \exp[2{x^{i}}(t)] g^{(i)} _{m_i , n_i}(y_i) dy^{m_i} _i dy^{n_i} _i
  \end{equation}
where the manifold $M_{i}$ with the metric $g^{(i)}$ is an
Einstein space of dimension $d_{i}$, $i = 1, \ldots ,n $; $n \geq
 2$. The matter sources had an equation of state with the pressures
being proportional to the densities
   \begin{equation} \label{1.3}
   {p_{i}^{(\nu)}} = \left(1-
   \frac{u_{i}^{(\nu)}}{d_{i}}\right) {\rho^{(\nu)}},
   \end{equation}
where $u_{i}^{(\nu)} = const$, with $i =1, \ldots ,n$ and  $\nu = 1,
 \ldots , m$. It was also assumed that
 \begin{equation}
  \label{1.4a}
    \quad \sum_{i=1}^{n} u_i ^{(\nu)}  > 0.
  \end{equation}
for all $\nu$. The densities were
restricted to be positive definite, ${\rho^{(\nu)}} > 0$, for any
component with a positive  square of the vector $u^{(\nu)}$ defined with
respect to the dual minisupermetric as
 \begin{equation}
  \label{1.4b}  \quad (u^{(\nu)})^2 =
   \sum_{i=1}^{n} \frac{(u^{(\nu)}_i)^2}{d_i} +
  \frac{1}{2 - D} \left( \sum_{i=1}^{n} u^{(\nu)}_i \right) ^2 > 0,
  \end{equation}
where $D$ is total dimension of $M$ in \eqref{1.1}.

Using the billiard approach \cite{IKM1,IKM} it was shown
in \cite{IMb0} that the solutions to higher dimensional Hilbert-Einstein
equations with pressures and densities satisfying the above conditions
had asymptotical Kasner-like behavior as $\tau \to + 0$
  \begin{equation}
   \label{1.5}
   ds^2_{as} = - d \tau ^2 +
   \sum_{i=1}^{n} A_i \tau ^{2 \alpha^i} 
   g^{(i)} _{m_i , n_i}(y_i) dy^{m_i} _i dy^{n_i} _i,
  \end{equation}
when there existed Kasner parameters in \eqref{1.5} obeying the
following two relations:
  \begin{equation}
  \label{1.6} \sum_{i=1}^{n} d_i \alpha^i  = 1, 
  \end{equation}
  \begin{equation}
  \label{1.7} 
  \sum_{i=1}^{n} d_i (\alpha^i)^2 = 1
  \end{equation}
and the following inequalities were valid
  \begin{equation}
   \label{1.8}
   \sum_{i=1}^{n} u_i^{(\nu)} \alpha^i   > 0 ~,~ 
   {\rm if} \qquad (u^{(\nu )})^2  > 0.
    \end{equation}
Here $\tau$ is the proper time given by $d \tau ^2 =
e^{2 \gamma (t)} dt^2$ and all $A_i > 0$.
When the condition \eqref{1.8} was not satisfied,
namely, if for {\bf any} Kasner set of parameters $\alpha$ there existed 
a fluid component  with index $\nu_0$ (obeying $(u^{(\nu_0)})^2 > 0$ ), 
such that 
 \begin{equation} \label{1.8a} 
 \sum_{i=1}^{n} u_i^{(\nu_0)} \alpha^i \leq 0, 
 \end{equation} 
one finds never-ending oscillating behavior near the singularity
similar to that of Bianchi-IX model. 

In \cite{IMb1} the above analysis was repeated but with the matter sources
given by scalar fields and form fields. The action for this
system was given by
  \begin{equation} \label{1.9}
    S_g=\int d^Dx
      \sqrt{|g|}\left( R[g]-h_{\alpha\beta}g^{MN}\partial_M \varphi^\alpha
    \partial_N \varphi^\beta-\sum_{a\in\Delta}\frac{\theta_a}{n_a!}
   \exp[2\lambda_a(\varphi)](F^a)^2 \right)
  \end{equation}
 where $g$ is the multidimensional metric with the line element
 $g_{MN}(x)dx^M dx^N$,
 $\varphi=(\varphi^\alpha)\in {\bf R}^l$, is vector of scalar fields, and
 $F^a =  dA^a$ is an $n_a$-form field. Also in \eqref{1.9}
 $h_{\alpha\beta}$ is a  constant, symmetric, positive definite,
 $l\times l$ matrix, $n_a\ge1$, $\theta_a=\pm1$,
 $\lambda_a(\varphi)=\lambda_{a \alpha} \varphi^\alpha$ gives the
coupling between the dilaton and form fields, $a\in\Delta$ with
 $\Delta$ being some finite set, and $\alpha=1,\dots,l$. The set $\Delta$ 
can be thought of as a set of ``color'' indices.

 In \cite {IMb1} the form fields in \eqref{1.9} were taken to have an 
 electric, composite brane ansatz, {\it e.g.} describing $S$-branes
 (or space-like branes). 
Previous, related work on cosmological and $S$-brane solutions can be 
found in \cite{BF}-\cite{IvS} and references therein.  For a $(p+2)$-form 
field this composite ansatz is written as $F = \sum_{I \in \Omega_{e}} d 
\Phi^{I} \wedge \tau(I)$ where $\Phi ^I$ is a scalar potential, $\Omega_e$ 
is the set of multi-indices of the form $\{ i_0, i_1, ..., i_p \}$ with 
$i_k$ being a number in the set ($1, ..., n$), and $\tau(I)$ is a $p+1$ 
volume form.  A non-composite ansatz would involve only a single term with 
no sum {\it i.e.} $F = d \Phi^{I} \wedge \tau(I)$. More details on the 
electric, composite ansatz are given in the next section.  Using this 
composite, electric ansatz some special classes of cosmological solutions 
where found on the manifold \eqref{1.1} and a billiard representation near 
the singularity was obtained, when certain restrictions on the brane 
configuration were imposed -- these included restrictions on brane 
intersections that guaranteed the block-diagonality of the stress 
energy-tensor (see also \cite{IMC}). Using  this representation it was 
shown that under certain conditions the solutions under consideration had 
a metric in \eqref{1.2} with asymptotic Kasner behavior as in \eqref{1.5} 
and with the scalar fields of the asymptotic form
  \begin{equation} \label{1.10}
   \varphi^\beta_{as}  = \alpha^{\beta} \ln \tau + const,
  \end{equation}
as $\tau \to +0$. The set of Kasner parameters for the metric
and scalar fields,
  $\alpha = (\alpha^A) = (\alpha^{i}, \alpha^{\beta})$,
obeyed \eqref{1.6} and the following relation
  \begin{equation} \label{1.11}
  \sum_{i=1}^{n} d_i
  (\alpha^i)^2 + \alpha^{\beta} \alpha^{\gamma}
  h_{\beta \gamma}= 1 ~
  \end{equation}
instead of \eqref{1.7}. The conditions under which this asymptotic Kasner
behavior of the metric and scalar field occurred were formulated
in terms of brane $U^s$ co-vectors
 \begin{equation} \label{1.12}
    U^s(\alpha) \equiv  U^s_A \alpha ^A 
    \equiv \sum_{i \in I_s} d_i \alpha^i
    -\chi_s\lambda_{a_s \gamma}\alpha^{\gamma},
  \end{equation}
with the scalar products  being positive definite
 \begin{equation} \label{1.13}
  (U^s,U^s) = d_s \left( 1 + \frac{d_s}{2-D} \right)+
 \lambda_{a_s \alpha} \lambda_{a_{s} \beta} h^{\alpha \beta} > 0.
 \end{equation}
Namely, one obtained asymptotic Kasner type behavior if for
all branes  (with $(U^s,U^s) > 0$)
  \begin{equation} \label{1.13a}
   U^s (\alpha) >0.
  \end{equation}

In the above $d_s$ is the dimension of the brane world-volume
manifold, $a_s \in \Delta$ is a color index for the brane, 
and $\chi_s  =   +1, -1$ for
electric or magnetic brane, respectively. Here for any brane
labeled by index $s$ the set $I_s =   \{ i_1, \ldots, i_k \}$,
 $i_1 < \ldots < i_k$, describes the location of brane, {\it i.e.}
the brane is ``attached'' to the sub-manifold  $M_s = M_{i_1}
 \times  \ldots \times M_{i_k}$. The vectors $u^{(\nu)} _i$ and 
 $U^s _A$ play similar roles for the respective cases of perfect fluid sources 
 and brane sources. In the case when the condition 
 \eqref{1.13a} was not satisfied, namely, if for {\bf any} Kasner
set of parameters, $\alpha = (\alpha^A)$, there existed a
brane with an index $s_0$ (obeying $(U^{s_0},U^{s_0}) > 0$), such
that
  \begin{equation} \label{1.14}
    U^{s_0}(\alpha) \leq 0,
  \end{equation}
one obtained never-ending oscillating behavior near the singularity.

It should be stressed that the billiard approach of \cite{IMb1}
was derived in a class of block-diagonal metrics defined on the
product of Einstein spaces \eqref{1.1} with certain restrictions imposed
on the signatures of the Einstein metrics, $g^{(i)}$. These
restrictions  are similar to the conditions $\rho^{(\nu)} > 0$
for $(u^{(\nu)})^2 > 0$ in the multicomponent fluid case.

In \cite{DH} Damour and Henneaux conjectured that the never-ending
oscillating behavior is generic to a certain class of string
theory inspired cosmologies (see also \cite{DHN} for a review).
Moreover, the authors of \cite{DH} give a sketch of a proof of
this conjecture. Although the ansatz for the cosmological metric
in \cite{DH} contained $d$-beins leading to additional
gravitational walls in the billiards and the Kasner parameters
also depended of spatial coordinates, the analysis of \cite{DH},
in essence, deals with Bianchi-I type model with a maximal number
of composite electric branes that form the billiard walls
according to the prescription of \cite{IMb1}. The gravitational
walls are irrelevant for the proof in  \cite{DH}. The key point of
this proof is the verification of inequalities \eqref{1.14} for
the maximal set of electric branes. (In \cite{DH} the coefficients
 $U^s_A$ in the linear functions $U^s(\alpha) = U^s_A \alpha^A$
were called electric and magnetic exponents.) This verification
was carried out for $D =11$ supergravity ($M$-theory), and also
for the four $D = 10$ string theories -- types $II A$, $II B$, $I$
and heterotic ($H$) -- using dualities.

In their paper \cite{DH} the authors  did not sketch a proof of
their conjecture for pure cosmological non-diagonal metric with the line
element of the form
  \begin{equation} \label{1.15}
    ds^2 = - \exp{(2 \gamma(t))} dt^2 + g_{ij}(t) dy^i dy^j.
  \end{equation}
In essence the authors in \cite{DH} dealt  with
the pure cosmological diagonal Bianchi-I type  metric (the
non-homogeneity of the metric and the presence of $(D-1)$-beins
were irrelevant for the proof) but they did not 
take into consideration the $(D-1)(D-2)/2$ equations coming from
the non-diagonal part of Hilbert-Einstein equations.

This naturally leads to the question as to whether the conjecture
of Damour and Henneaux may be reformulated somehow
for composite  $S$-brane solutions
with diagonal  metrics. At present we have no examples of
solutions in string inspired cosmologies with never-ending
oscillating behavior near the singularity when the metric is
diagonal.

In this paper we point out that a certain class of composite
$S$-brane solutions with diagonal metrics can avoid this
oscillating behavior near the singularity even if the diagonal part of
Hilbert-Einstein equations leads to ``confining'' billiards.

In section 2 we give the general set up of our system of $(D=n+1)$-dimensional
gravity with $(p+2)$-form field.  We consider a composite $Sp$-brane ansatz
for the antisymmetric  $(p+2)$-form field, with all the ansatz
functions depending only on the time coordinate.
An important feature of the composite ansatz is that the
charge densities of the branes obey certain quadratic constraints
 \cite{IMC,IvS}. In \cite{IvS} it was found that these constraints
have ``maximal" solutions with all non-zero electric brane charge densities
in specific odd dimensions with specific rank of the form fields
({\it i.e.} $D=  5, 9, 13, \dots$ and $p = 1,3,5 \dots$, respectively).
In section 2 we review these solutions for the model of
 \cite{IvS} but without the scalar field.

In section 3 we give two explicit examples how the oscillating behavior near the
singularity  is stopped when one replaces non-composite $S$-branes
(non-composite usually means that for one form there are no more than
one brane) with the composite $S$-branes. The two examples both have diagonal metrics, and
spacetime dimension and $(p+2)$-forms given by $D=4$, $p=0$ and
$D=5$, $p=1$, respectively.

  %%%%%%%%%%%%%%%%%%%%%%%%%%%%%%%%%%%%%%%%%%%%%%%%%%%%%%%%%%%%%%%%%%%%%%%%
  \section{\bf D-dimensional gravity with $(p+2)$-form}
  %%%%%%%%%%%%%%%%%%%%%%%%%%%%%%%%%%%%%%%%%%%%%%%%%%%%%%%%%%%%%%%%%%%%%%%%

The multidimensional system considered here is similar to that of
\cite{IvS} but without the scalar field. The action is given by
  \begin{equation} \label{2.1}
   S =
      \int_{M} d^{D}z \sqrt{|g|} \left[ {R}[g]
   -  \frac{1}{q!}  F_{[q]}^2 \right],
  \end{equation}
where $g$ is the metric with the line element
$g_{MN} dz^{M} dz^{N}$ ($M,N =1, \ldots, D$), and
  \begin{equation} \label{2.2}
  F_{[q]} =  dA_{[q-1]} =
  \frac{1}{q!} F_{M_1 \ldots M_{q}}
  dz^{M_1} \wedge \ldots \wedge dz^{M_{q}} ,
  \end{equation}
is a $q$-form ($q =  p +2 \geq 2$) on a $D$-dimensional
manifold $M$. The manifold is of the form $ M = (t_{-}, t_{+})
\times {\bf R}^{n}$ with the first coordinate $t$ being the
distinguished, time-like coordinate. The metric is diagonal and of
the form
  \begin{equation} \label{2.3}
   ds^2 = - \mbox{\rm e}^{2{\gamma}(t)} dt^2 +
   \sum_{i= 1}^{n} \mbox{\rm e}^{2 \phi^i(t)}  (dy^i) ^2 .
  \end{equation}
The ansatz functions $\gamma (t), \phi ^i (t)$,
and the $q$-forms are assumed to depend only on the distinguished,
time-like coordinate $t$. The functions $\gamma,\phi^i$ are smooth.

The $(q = p+2)$-form, $F$, is taken in the electric
composite $Sp$-brane ansatz of the form
  \begin{equation} \label{2.4}
      F = \sum_{I \in \Omega_{e}} d \Phi^{I} \wedge \tau(I)
  \end{equation}
where $\Phi ^I$ is a scalar potential. This ansatz is composite
since it is a sum of monomial terms rather than a single term for
the non-composite case. In
\eqref{2.4}  $\Omega_e$ is the set of multi-indices having the
form $\{ i_0, i_1, ..., i_p \}$ (all  $i_k$ are among the numbers
$1, ..., n$).

In \eqref{2.4} we have introduced a volume form of rank
$d(I) = p+1$
  \begin{equation} \label{2.6}
  \tau(I) \equiv dy^{i_0}  \wedge \ldots \wedge dy^{i_p},
  \end{equation}
here $ I = \{ i_0, ..., i_p \}$ which corresponds to the brane
submanifold with the coordinates $y^{i_0},   \ldots,  y^{i_p}$.

The scalar potential depends only on the distinguished, time-like
coordinate, {\it i.e.} $\Phi^I = \Phi^I(t)$.
The solution of the ``Maxwell" equations for the scalar potentials
$\Phi^I (t)$ has the following form \cite{IvS}
  \begin{equation} \label{2.6a}
   \dot \Phi ^I=Q_I \exp(2U^I), \qquad
    U^I = U^I(\phi)= \sum_{i \in I} \phi^i.
  \end{equation}
where $Q_I$ are constant charge densities and $I \in \Omega_{e}$.
The $q=(p+2)$-form  then reads
   \begin{equation} \label{2.6b}
    F_{[q]} =\frac{1}{(p+1)!} Q_{i_0  \dots i_p}
      \exp(2 \phi^{i_0} + \dots + 2 \phi^{i_p} )
      du \wedge d y^{i_0} \wedge \dots \wedge d y^{i_p}
  \end{equation}
where $Q_{i_0  \dots i_p}$ are components of antisymmetric
form coinciding with $Q(I)$ for $i_0 < \dots < i_p$.

Since the Ricci tensor is diagonal one finds
  \begin{equation} \label{2.6c}
      F_{i M_2 \dots M_{q}} F_{j}^{M_2 \dots M_q}
  \propto T_{i j} = 0, \qquad i \neq j.
  \end{equation}
where $T_{ij}$ is the stress-energy tensor. This leads
to constraints among the charge densities, $Q_{i_0  \dots i_p}$. 
Absorbing the functional dependence on $t$ into the definition of the
charge densities as
   $\bar{Q}_{i_0  \dots  i_p}$ via
   \begin{equation} \label{2.6e}
    \bar{Q}_{i_0  \dots  i_p} \equiv  Q_{i_0  \dots i_p}
    \prod_{k = 0}^{p} \exp(\phi^{i_k}) ~,
   \end{equation}
one can write the constraints as \cite{IvS}
   \begin{equation} \label{2.6d}
  C_i ^j \equiv \sum_{i_1, \dots, i_p =1}^{n}
    \bar{Q}_{i i_1 \dots i_p} \bar{Q}^{j i_1 \dots i_p} = 0,
   \end{equation}
$i \neq j$, $i, j =1, \dots,n$. Here the indices are lifted by 
the flat Euclidean
metric $(\eta _{ab}) = (\eta ^{ab}) = diag(1 , \dots , 1)$.
These $\bar{Q}_{i_0  \dots i_p}$ can be viewed as ``running"
charge densities with the functional dependence coming from
$\phi^{i_k} (t)$.

In \cite{IvS} it was shown that for
    \begin{equation} \label{2.d}
     D = 4m +1, \qquad p = 2m-1,
    \end{equation}
these constraints could  be satisfied if the non-running charge densities 
were self-dual or anti-self-dual in a flat Euclidean space ${\bf R}^n$, 
   {\it i.e.}
    \begin{equation} \label{2.6f}
     Q_{i_0  \dots i_p} =
    \pm \frac{1}{(p+1)!} \varepsilon _{i_0  \dots i_p j_0 \dots j_p}
     Q^{j_0 \dots j_p} = \pm (* Q)_{i_0  \dots i_p} ~,
    \end{equation}
and if all the scale factors were the same
 \begin{equation} \label{2.6g}
   \phi ^i (t)  = \phi (t).
 \end{equation}
In \eqref{2.6f} 
$\varepsilon _{i_0  \dots i_p j_0 \dots j_p}$ is the completely
anti-symmetric symbol.

Along the lines of \cite{IvS} (when a scalar field was present)
the equations of motion for the action \eqref{2.1} have
the following solution when relations \eqref{2.d} are
satisfied:
   \begin{eqnarray} \label{2.8}
   \phi (t) &=& \frac{2}{n} f(t)~, 
   \qquad f(t) = - \ln \left[ (t-t_0) |Q^2 K|^{1/2} \right], \\
    \label{2.8b}
     ds^2 &=& -e^{2n \phi(t)} dt^2 
     + e^{2 \phi(t)} \sum _{i=1}^n (dy^i)^2, \\
    \label{2.8c}
    F &=& e^{2 f(t)} dt \wedge Q,
    \qquad
    Q = \frac{1}{(p+1)!} Q_{i_0  \dots i_p}
                          dy^{i_0}  \wedge \dots \wedge dy^{i_p},
    \end{eqnarray}
where $K = - \frac{n}{4(n-1)}$ and $Q^2 = \sum_{I \in \Omega_e}
 Q^2(I)$. In the above the ansatz function $\gamma (t) = \sum_{i=1}
 ^n \phi ^i (t) = n \phi (t)$ because of \eqref{2.6g}. The charge
density form  $Q$ is of rank $n/2 = 2m$ and is self-dual or
anti-self-dual in a flat Euclidean space ${\bf R}^n$:  $Q = \pm *
Q$. This solution may be obtained from that of reference
\cite{IvS} by taking the parameters as $C = C_1 = C_2 = \lambda =
0$. Note that this solution with composite form field does not
have oscillating Kasner like behavior as one approaches the
singularity at $t \rightarrow + \infty$ (or $\tau \to + 0$).

It should be stressed that in the case $D =5$ and $p =1$, the
solution presented above is a general one when all charge
densities $Q(I)$ are non-zero, $I \in \Omega_e$. More, rigorously,
here as in \cite{IvS}, we get that the running ``charge densities"
$\bar{Q}_{i_0 i_1}$ are self-dual or anti-self-dual and all  scale
factors are the same up to constants
     \begin{equation} \label{4.ba}
      \phi^i(t) = \phi(t) + c^i,
     \end{equation}
 $i = 1,...,4$. The constants $c^i$ may be absorbed by an
appropriate rescaling of $y^i$ coordinates.

 \section{Breaking of  oscillating behavior near
 the singularity by constraints}

In  \cite{IMb1} it was found that certain cosmological models with
$p$-branes may have a ``never ending" oscillating behavior, {\it i.e.}
as one approaches the cosmological singularity the Kasner parameters
of the asymptotic metric ``jump'' between different locally constant values.
This  is similar to the behavior found in Bianchi-IX model \cite{BKL} and
in multidimensional models with multicomponent perfect fluid 
\cite{IKM1,IKM,IMb0}.

This oscillating behavior may be described graphically using the
so-called billiard representation near the singularity. The point
described by the anisotropic part of the minisuperspace
coordinates $z = (z^a)$ ( $a = 1,..., n-1$) moves in the asymptotical
regime near the singularity in a region enclosed by ``hard walls"
corresponding to inequalities on the Kasner parameters
such as \eqref{1.13a}.  The $z^a$ are linear combinations of the 
$x^i (t)$ from \eqref{1.2}, {\it i.e.} they 
are combinations of logarithms of the scale factors. The point $z(t)$
traces out path in this region until it encounter a ``wall" at
which point it recoils and moves toward another ``wall". An
example of the billiard approach for a model with scalar fields
and form fields (with the form fields having a composite $S$-brane
ansatz) can be found in \cite{IMb1}.

Here we show that this oscillating behavior near the singularity
can be broken in certain cases due to the compositeness of the branes
and the diagonality of the metric. We show this via two examples:
$D=4$ with $2$-forms, and $D=5$ with $3$-forms. In each case the
breaking of the oscillating behavior is a result of the
constraints on charge densities, following from the non-diagonal part
of Hilbert-Einstein equations.

  \subsection{$4$-dimensional  model with $2$-form}

Let us consider the $4$-dimensional model
   \begin{equation} \label{3.1}
      S_{4} =
            \int_{M} d^{4} z \sqrt{|g|} \left[ {R}[g] -
            \frac{1}{2!}  F^2_{[2]} \right],
   \end{equation}
with the metric $g$ and $2$-form $F_{[2]} =  dA_{[1]}$ defined on
a $4$-manifold  $M$. The metric is taken as diagonal and of the form
    \begin{equation} \label{3.2}
      ds^2 = - \mbox{\rm e}^{2{\gamma}(t)} dt^2 +
          \sum_{i= 1}^{3} \mbox{\rm e}^{2 \phi^i(t)}  (dy^i) ^2, \qquad
          {\rm with} \qquad \gamma  =  \sum_{i= 1}^{3} \phi^i.
    \end{equation}
The electric $2$-form has the composite form
    \begin{equation} \label{3.3}
      F_{[2]} = \sum_{i = 1}^{3} d \Phi^i \wedge dy^i.
    \end{equation}
From  \eqref{2.6a} we get
    \begin{equation} \label{3.4}
      \dot\Phi^i = Q_i \exp(2 \phi^i),
    \end{equation}
$i = 1,2,3$.

For this case the constraints from \eqref{2.6d} are
    \begin{equation} \label{3.5}
     C_{ij} =  Q_{i} Q_{j } = 0,
    \end{equation}
$i \neq j$; $i,j = 1,2,3$.  It follows from  \eqref{3.5} that only one 
charge  $Q_{i_0}$, may be non-zero. At the end of this section we will use the billiard
representation to show that this implies one has asymptotic Kasner behavior, rather
than never-ending oscillating behavior near the singularity.

Now let us omit for a moment the constraints \eqref{3.5}, {\it i.e.}
the non-diagonal part of Einstein equations is not considered.
Such a ``truncation" is equivalent to considering a model with
three $2$-forms, having the following action
   \begin{equation} \label{3.6}
    S_{4br} =
      \int_{M} d^{4} z \sqrt{|g|} \left[ {R}[g] -
     \frac{1}{2!} \sum_{i=1}^{3} (F^i_{[2]})^2 \right],
   \end{equation}
instead of \eqref{3.1} and with a non-composite ansatz
     \begin{equation} \label{7.4ba}
      F^i_{[2]} =  d \Phi^i \wedge dy^i,
    \end{equation}
$i = 1,2,3$, instead of \eqref{3.3}. Relation \eqref{3.4} survives
in this non-composite case.

The asymptotical behavior near the singularity ($\tau \to + 0$) 
for this truncated or
non-composite model with  non-zero charges is described by the
metric \cite{IMb1}
      \begin{equation} \label{3.7}
       ds^2 _{as} = - d\tau ^2
                + \sum_{i=1}^{n} A_i \tau^{2 \alpha^i(\tau)} (dy^i) ^2,
       \end{equation}
with the Kasner parameters satisfying the relations
       \begin{equation} \label{3.7a}
       \sum_{i=1}^{n}  \alpha^i =
           \sum_{i=1}^{n} (\alpha^i)^2 = 1,
       \end{equation}
$A_i > 0$  are constants and $n=3$. The Kasner ``parameters"
$\alpha^i(\tau)$ are  constant in the intervals $\tau \in [\tau_{k
+1}, \tau_{k})$, $k = 0, 1, 2, \dots$,  and $\tau_k \to +0$ as $k
\to \infty$, but take different values for neighboring intervals.

Thus, one gets never-ending oscillating behavior  near the
singularity, equivalent to that of the Bianchi-IX model
\cite{BKL}. The dynamics of this ``motion" may be described
graphically as the motion of a ``particle" in the well-known
triangle billiard ``table" belonging to Lobachevsky space
\cite{Ch} (see also \cite{IMb0,Pull,Kir1,Kir2}). The reflection from
a wall describes the change to a different Kasner ``epoch''. The
``collision law" for the ``particle"  in terms
of Kasner parameters may be obtained from general relations
given in \cite{DH,Ierice}. This triangular billiard ``table" is
depicted in figure 1. It is the interior triangle with curved, hard
walls labled by $+\infty$. These infinite positive walls result from 
the fact that  the scalar product of all $U$-vectors obey
       \begin{equation} \label{3.8}
         (U^I,U^I) \equiv G^{ij} U^I_i U^I_j = \frac{1}{2} > 0,
       \end{equation}
for $I = \{i \}$, $i = 1, 2, 3$, \cite{IMb0,IMb1}. 
Relation \eqref{3.8} is obtained from \eqref{1.13}
with $d_s =1$  and $h^{\alpha \beta} =0$. Here and in the
next subsection $G^{ij} = \delta_{ij} + \frac{1}{2-D}$ are the
components of the inverse matrix to the matrix of the
minisuperspace metric $G_{ij}= \delta_{ij}- 1$ with $i,j= 1,\dots,n$
(for its diagonalization, see for example, \cite{IMZ}).
The enclosure of the billiard table in figure 1 is not compact 
but it has a finite area.

\begin{figure}
\includegraphics[bb=0 10 525 515,height=6cm,width=8cm]{fig1(singleton).ps}
\caption{The triangle billiard for $D=4$, 
$q=2$. The hard walls, $+\infty$, come from \eqref{3.8}. The 
``particle" described by the minisuperspace coordinates, $z^a$, moves in 
this billiard until it encounters a wall and is reflected. The 
enclosure of the billiard is non-compact but the area is finite.}
\end{figure}

The never-ending oscillating behavior for the non-composite model
takes place since for any Kasner set $\alpha = (\alpha^i)$ there exists
a brane $U$-vector satisfying
       \begin{equation} \label{3.9}
            U^{I}(\alpha) = \alpha^{i} \leq 0,
       \end{equation}
for $I = \{i \}$, $i = 1, 2, 3$, \cite{IMb1}. We note  that
the equality in the right hand side of \eqref{3.9}
takes place at the three vertices of the triangle, namely, $\alpha
= (0,0,1), (0,1,0), (1,0,0)$.

Returning to the composite model, the constraints \eqref{3.5}
imply that two of the charges vanish and, hence two of the walls
of the billiard ``table" vanish. This opens a region at infinity
for a non-oscillating asymptotical Kasner type
behavior. 

This example has a straightforward generalization
to $(n+1)$-dimensions. It is interesting to note that
as it was shown recently in \cite{BKM} in the general case
of non-diagonal metric \eqref{1.15} the oscillating behavior
near the singularity is recovered for  the $(n+1)$-dimensional
model with a $2$-form. In this case there is (asymptotically) one
``jumping'' wall that effectively acts as a full set of walls.

  \subsection{$D = 5$ model with $3$-form}

Now we consider the following $5$-dimensional model
   \begin{equation} \label{3.10}
    S_{5} =
      \int_{M} d^{5} z \sqrt{|g|} \left[ {R}[g] -
     \frac{1}{3!}  (F_{[3]})^2 \right],
   \end{equation}
with metric $g$ and $3$-form   $F_{[3]} =  dA_{[2]}$ defined on a
$5$-dimensional manifold $M$.

It follows from the results of section 2 that the general solutions
with maximal number of
composite electric $S1$-branes (strings)  resulting from the action
\eqref{3.10} with a diagonal metric are
    \begin{equation} \label{3.11}
      ds^2 = - \mbox{\rm e}^{8{\gamma}(t)} dt ^2 +
          \sum_{i= 1}^{4} \mbox{\rm e}^{2 \phi(t) + 2 c^i}  (dy^i) ^2,
    \end{equation}
and
    \begin{equation} \label{3.12}
      F_{[3]} = \sum_{I} d \Phi^{I} \wedge \tau(I),
    \end{equation}
where $I = \{i,j \}$, $i <j$; $i,j =1, 2, 3, 4$, and all charges
 $Q_I$ are non-zero. As can be seen from \eqref{2.8} - \eqref{2.8c}
this solution has neither an oscillating nor
even a Kasner-type behavior ({\it i.e.} writing the metric in Kasner form, 
\eqref{3.7}, the Kasner powers do not satisfy \eqref{3.7a})  near the singularity.

Now we omit the constraints \eqref{2.6d}. This leads
to a model with six $3$-forms having the action
    \begin{equation} \label{3.13}
      S_{5br} =
      \int_{M} d^{5} z \sqrt{|g|} \left[ {R}[g] -
      \frac{1}{3!} \sum_{I} (F^I_{[3]})^2  \right],
    \end{equation}
instead of \eqref{3.10} and a non-composite ansatz
     \begin{equation} \label{3.14}
      F^I_{[3]} =  d \Phi^I \wedge  \tau(I),
    \end{equation}
instead of the composite one \eqref{3.12}.

We can use the results of \cite{IMb1} to show that for these
non-composite solutions with non-zero charge densities, $Q_I$, of
the $1$-branes (strings) we get never-ending oscillating
behavior near the singularity as $\tau \to + 0$, described by the
metric formulas \eqref{3.7}, \eqref{3.7a} with $n =4$ and certain
locally constant functions $\alpha^i(\tau)$.

In this case we get a $3$-dimensional billiard ``table" with six
walls that form a deformed cube. From the relationships in
\cite{IMb1} the explicit form of this billiard 
belonging to the $3$-dimensional Lobachevsky space (``ball")
$H^3= D^3 = \{ \vec{z} \in {\bf R}^3 : |\vec{z}| < 1 \}$ is
given by the six inequalities
     \begin{equation} \label{3.bil}
      |\vec{z} - \vec{v}_k| > \sqrt{2},
    \end{equation}
  $k = 1,...,6$, where
     \begin{equation} \label{3.v}
    \vec{v}_1 = - \vec{v}_6 = (\sqrt{3} ,0,0), \quad
    \vec{v}_2 = - \vec{v}_5 = (0,\sqrt{3},0), \quad
    \vec{v}_3 = - \vec{v}_4 = (0,0,\sqrt{3}).
 \end{equation}

As in the previous example the enclosure of the billiard is
not compact, but the volume is finite. This can be seen in
terms of the ``illumination" of the Kasner sphere by  point
sources of ``light".  The sources of light located at the points
 $\vec{v}_k$, $k = 1,...,6$, completely illuminate the unit sphere
 $S^2 = \{ \vec{z} \in {\bf R}^3 |  |\vec{z}| = 1 \}$, and
according to the ``illumination'' theorem of \cite{IMb0,IMb1} the
billiard has a finite volume.

As in the previous example the walls come from the fact that the scalar
products of all $U$-vectors obey
       \begin{equation} \label{3.15}
            (U^I,U^I) \equiv G^{ij} U^I_i U^I_j = \frac{2}{3} > 0,
       \end{equation}
for $I = \{i, j \}$, $i < j$, \cite{IMb0,IMb1}. In \eqref{3.15} we have
used \eqref{1.13} with $d_s=2$ and $h^{\alpha \beta} = 0$.

One will get never-ending oscillating behavior for the
non-composite model since for any Kasner
set, $\alpha = (\alpha^i)$, there exists
a brane $U$-vector which satisfies
       \begin{equation} \label{3.16}
           U^{I}(\alpha) = \alpha^{i} + \alpha^{j} \leq 0,
       \end{equation}
for $I = \{i, j \}$, $i < j$. We note that the
equality in the right hand side of \eqref{3.16} takes place for
$\alpha = (0,0,0,1), (\frac{1}{2}, \frac{1}{2}, \frac{1}{2}, -
\frac{1}{2})$ and $6$ other sets obtained by permutations. These are points 
on the Kasner sphere $S^2$ coinciding with the eight  vertices of the
``deformed" cube.

In the composite case, as we have seen from the exact solution
given by \eqref{2.8} - \eqref{2.8c},
there is no oscillating behavior of the scale factors. From the
billiard representation point of view this solution corresponds to
a static point in the billiard.

\section{Conclusions}

In this paper we have presented examples showing that the
never-ending oscillating behavior near the singularity with
diagonal metric can be broken when non-composite branes are
replaced by composite ones.

The mechanisms of breaking the oscillating behavior are different 
in these examples.
In the $4D$ case the ``constraints''  \eqref{2.6d}
destroy all the walls of the billiard
``table" except one. In the $5D$ case in general some of the walls could also be
``destroyed'' due to zero charge densities. But in our example all walls
survive since all the charge densities $Q_{ij}$ are non-zero and obey
(anti)-self-duality relations. In this case the constraints select
only those solutions that have  simple isotropic behavior (with
non-Kasner and non-oscillating behavior near the singularity).

These examples suggest a conjecture that an analogous effect may
take place for $11$-dimensional supergravity and $10$-dimensional
models of superstring origin. In other words, one may consider the
following hypothesis: for certain supergravity models (say, for
$D=11$, or $D =10$ $II A$ supergravities) there are no composite
$S$-brane solutions with (block)-diagonal metrics that have a
never-ending oscillating behavior near the singularity.

We note that in \cite{IMb1} an example of oscillating behavior
described by a $4$-dimensional billiard  of finite volume in
``truncated'' $D = 11$ supergravity model (without the
Chern-Simons term) was suggested. This billiard  was supported by
ten magnetic $5$-branes.  It was shown that the inclusion of the
Chern-Simons term destroyed some walls of the billiard ``table"
and allowed an asymptotical Kasner-like behavior as $\tau \to
+0$. Interestingly, the constraints of Chern-Simons origin in
 \cite{IMb1} look very similar to the constraints considered in
this paper. It should be pointed out that recently some other mechanisms for avoiding 
oscillating behavior near the singularity were suggested in 
 \cite{Wesley1,Wesley2}.

The main conclusion of this paper is that the  never-ending
oscillating behavior near the singularity of solutions in certain
cosmological models with forms (and scalar fields) can be
broken or avoided. This avoidance of the  oscillating behavior can
be attributed to the form of the ansatz considered here: a diagonal
metric, and composite electric ansatz for the form-fields. This result
comes from the constraints on the brane charge densities which are due
to the diagonality of the metric and the compositeness of the brane
system.  This shows that such behavior may not be generic to such
higher dimensional cosmologies.

 \begin{center}
 {\bf Acknowledgments}
 \end{center}

 The work of V.D.I. and V.N.M.was supported in part by a DFG grant
 Nr. 436 RUS 113/807/0-1 and by the Russian Foundation for
 Basic Researchs, grant Nr. 05-02-17478. D.S. would like
to thank Prof. Max Chaves for the invitation to work at the Universidad
de Costa Rica.

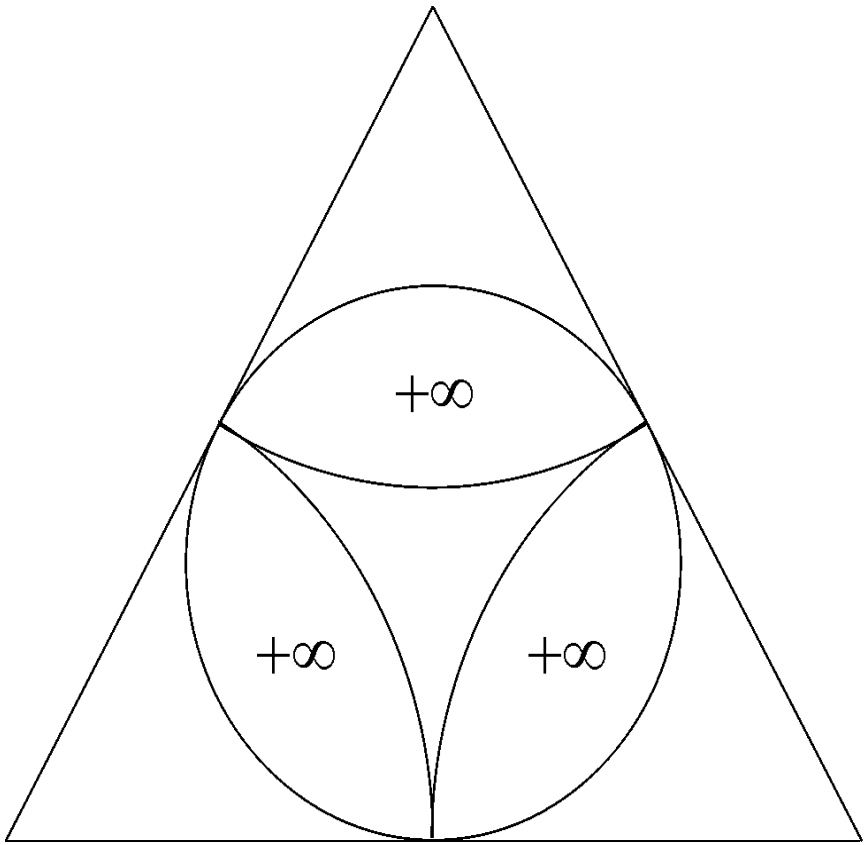
\end{document}